# Characteristic earthquake model, 1884 – 2011, R.I.P.

Yan. Y. Kagan (kagan@moho.ess.ucla.edu)
ESS/UCLA, Los Angeles, CA 90095-1567, USA

David D. Jackson (david.d.jackson@ucla.edu)
ESS/UCLA, Los Angeles, CA 90095-1567, USA

Robert J. Geller (bob@eps.s.u-tokyo.ac.jp)
Department of Earth and Planetary Science
University of Tokyo, Tokyo 113-0033 Japan

ABSTRACT. Unfortunately, working scientists sometimes reflexively continue to use "buzz phrases" grounded in once prevalent paradigms that have been subsequently refuted. This can impede both earthquake research and hazard mitigation. Well-worn seismological buzz phrases include "earthquake cycle," "seismic cycle," "seismic gap," and "characteristic earthquake," recorded in the ISI Web of Science data base for the period 2009-2012 66, 88, 84, and 22 times, respectively. The primary assumption loading these phrases is that there are sequences of earthquakes that are nearly identical except for the times of their occurrence. If so, the complex process of earthquake occurrence could be reduced to a description of one "characteristic" earthquake plus the times of the others in the sequence. The "seismic gap" (or the effectively equivalent "seismic cycle") model depends entirely on the "characteristic" assumption. The gap models assume quasi-periodic behavior of something, and that something must be characteristic earthquakes. However, since 1990s numerous statistical tests have failed to support characteristic earthquake and seismic gap/cycle models, and the 2004 Sumatra earthquake and 2011 Tohoku earthquake both ripped through several boundaries between supposed segments. Earthquake scientists must take a rigorous look at characteristic earthquake models and their implicit assumptions, and should scrap merely long-standing ideas that have been rejected by objective testing or are too vague to be testable.

A precept of science is that theories unsupported by observations and experiments must be corrected or rejected, however intuitively appealing they might be. Unfortunately, working scientists sometimes reflexively continue to use "buzz phrases" grounded in once prevalent paradigms that have been subsequently refuted. This can impede both earthquake research and hazard mitigation.

Well-worn seismological buzz phrases include "earthquake cycle" (66 instances recorded in the ISI Web of Science data base for the period 2009-2012), "seismic cycle" (88 instances), "seismic gap" (84), and "characteristic earthquake" (22). The grand prize goes to the trifecta of "fault," "segment," and "earthquake" with 546 hits. Each phrase carries heavy baggage of implicit assumptions. The primary assumption loading these phrases is that there are sequences of earthquakes that are nearly identical except for the times of their occurrence. If so, the complex process of earthquake occurrence could be reduced to a description of one "characteristic" earthquake plus the times of the others in the sequence. Often such a characteristic earthquake sequence is assumed to dominate the displacement on fault or plate boundary "segments." This view holds that characteristic earthquakes should be the largest on a given segment and exhibit quasi-periodic recurrence; it thus has characteristic earthquakes occurring at a rate higher than that implied by the classic Gutenberg-Richter distribution.

The problem is that the surmised properties of "characteristic earthquakes" were inferred by selecting examples from the past and have proven too imprecise to apply to future earthquakes. Perhaps the best-known example of a "characteristic" sequence is that near Parkfield, California, which was the basis for a 1985 prediction that there was a 95% probability of a "repeat" before 1993 (see Bakun et al., 2005, and its references). The example sequence included six events, of which only two were recorded by a California seismic network. There were many published descriptions of the hypothetical characteristic earthquake, but the only consistent features were "on the San Andreas fault," "near Parkfield," and "about magnitude 6." Much attention was paid to the fact that no qualifying event occurred before 2004 (eleven years after the end of the prediction window), but little was focused on the ambiguities of what was predicted. Any event with magnitude between 5.5 and 7.5 and rupture length over 20 km would arguably have satisfied at least some of the published descriptions.

Jordan (2006) pointed out that a scientifically valid hypothesis must be prospectively testable. Ironically, his article made the untestable assertion that "the northern San Andreas is entering a mature stage of the Reid cycle." The buzz phrases die hard. Retrospective analyses cannot provide a rigorous foundation for any model of earthquake occurrence including, but not limited to, the "seismic cycle." Even the simplest spatial window, a circle, has three degrees of freedom for its characterization. The famous mathematician and physicist John von Neumann remarked that with four parameters he could "fit an elephant..." (Dyson, 2004). Furthermore, retrospective searches of seismicity patterns can usually find seemingly significant features in completely random simulations (Shearer and Stark, 2012).

The case of Parkfield shows how retrospective analysis can mislead. The presumed characteristic earthquakes were selected from several different types of catalog without a clear set of guidelines. By contrast, Fig. 1 shows a magnitude distribution plot for cataloged instrumentally measured earthquakes within the first published polygon describing where the characteristic events should occur. The recorded earthquakes, including the late arriving 2004 magnitude 6 event, are fit well within the 95% confidence limits by a Gutenberg-Richter distribution. In contrast, the characteristic earthquake model would imply a surplus of magnitude 6 events exceeding the Gutenberg-Richter upper confidence limit.

The concept of seismic gaps was broached by Gilbert (1884) and Reid (1911) well before plate tectonics was proposed. Implicitly assuming that recent large earthquakes were characteristic (although he did not use that term), Fedotov (1965) postulated that segments (he did not use that term either) of plate boundaries (to use modern terminology) that had not ruptured for some time were due for large earthquakes. His hypothesis, if true, would have significantly advanced long-term forecasting.

The "seismic gap" (or the effectively equivalent "seismic cycle") model depends entirely on the "characteristic" assumption. The gap models assume quasi-periodic behavior of something, and that something must be characteristic earthquakes. By itself the phrase "characteristic" may describe spatial or size properties without implying quasi-periodicity, but "characteristic" explicitly or implicitly connotes quasi-periodicity if it is grouped with 'gap' or 'cycle.' As used in the gap model, the characteristic hypothesis often brings even more baggage: characteristic events are assumed to be the largest possible on a segment.

The seismic gap model has been used to forecast large earthquakes around the Pacific Rim. However, testing of these forecasts in the 1990s and later revealed that they performed worse than random Poisson forecasts (see Rong et al., 2003 and its references). Similarly, the characteristic earthquake model has not survived statistical testing (see Jackson and Kagan, 2011, and its references). Yet despite these clear negative results, the characteristic earthquake and seismic gap models continue to be invoked.

Some proponents of quasi-periodic characteristic earthquakes draw support from paleoseismic data which provide radiometric dates of sediments bracketing successive earthquake ruptures at specific trench sites along a fault. In some cases these data give additional information such as components of the displacement vector between the sides of a fault. But it is hard to draw firm conclusions about temporal regularities from such generally imprecise and irreproducible data. Why is this so? Some earthquakes may not displace the trench sites. Sample collection and analysis are subjective. Unless significant sedimentation occurs between successive earthquakes, they're indistinguishable. The data cannot determine magnitudes or other properties that could be used to tell characteristic earthquakes from others. Nevertheless, some measured sequences of dates appear quasi-periodic and inconsistent with Poisson behavior. Comprehensive studies (e.g., Parsons, 2008) suggest that earthquakes at some sites appear quasi-periodic, while others don't (e.g., Grant, 1996; Grant and Sieh, 1994). Rarely discussed is the fact that some short sequences drawn from a random process would appear quasi-periodic, others Poissonian, and others clustered. Also, multiple earthquakes separated by time intervals within the error bars could be interpreted as one event, biasing the interpretation towards quasi-periodic occurrence.

The unproven assumptions in the characteristic earthquake and seismic gap models are far from benign. They lead to overestimating the rates of characteristic magnitude earthquakes, often accompanied by underestimating the maximum size and rates of larger events not envisioned in the characteristic model. Clear examples of such "uncharacteristic" earthquakes include the disastrous 2004 Sumatra and 2011 Tohoku mega-quakes, each of which ripped through several previously hypothesized "segment boundaries." These tragic failures should have been the last straw. What additional evidence could possibly be required to refute the characteristic earthquake hypothesis? Worse yet: the gap model sends a false message of relative safety. It implies that in the aftermath of a characteristic earthquake a region is immune from further large shocks, yet comprehensive studies (e.g. Kagan and Jackson, 1999) show that large earthquakes increase the probability at all magnitudes.

Yet the hit counts above show how time-worn models can linger on. Their baggage of implicit assumptions often arrives undeclared and goes unquestioned. What can be done to overcome this intellectual inertia?

First, reviewers, editors, and funding agency officials must recognize that there is a problem. The buzzwords discussed above, like the words 'forecast' and 'prediction,' have been used quite loosely. Terms in publications and proposals should be defined clearly, evidence for assumed segmentation and characteristic behavior should be critically examined, dependence on and rules for selected data should be made explicit, and full prospective tests for unproven assumptions should be described.

Second, those responsible for hazard estimates should emphasize models that deal with the whole spectrum of earthquake occurrence, not characteristic and other models based on small subsets of arbitrarily selected data. Statistical studies based on *all earthquakes* within given time, space, and size limits–not just those which fit the investigator's preconceptions—are advancing rapidly and should become our new standard. All of us in earthquake science must wake up to the problems caused by relying on selected data. Arbitrarily chosen datasets are fine for *formulating* hypotheses, but not for *validating* them.

Third, characteristic advocates and statistical hypothesis testers should collaborate to develop appropriate tests. Collaboratory for the Study of Earthquake Predictability (CSEP) centers in California, Japan, Switzerland, and New Zealand are prospectively testing hundreds of regional and global earthquake models (see http://cseptesting.org/centers/scec). As far as we know, no characteristic earthquake or seismic cycle models are included, but they could be.

Nishenko (1991) and the coauthors of his earlier papers deserve enormous credit for articulating the seismic gap model in a testable form. Their specific models failed, but that's how science works. True believers should now either throw in the towel or reformulate the characteristic model for another explicit prospective test. A problem for such testing is that modern models often address limited regions in which definitive earthquakes may not occur for centuries. A model must be, like Nishenko's, basic enough to cover a region with a high combined rate of relevant events whose number will be sufficient (usually 15-20) for decisive testing within a few years

In summary, the time for case studies, anecdotes, speculation, and band-aids for failed models has passed. We must take a rigorous look at characteristic earthquake models and their implicit assumptions. In the play *Helen*, Euripedes wrote that "Man's most valuable trait is a judicious sense of what not to believe." Let's begin by scrapping merely long-standing ideas that have been rejected by objective testing or are too vague to be testable. Otherwise, earthquake science won't deserve the name.

Shearer, P., and Stark, P. B. (2012). Global risk of big earthquakes has not recently increased. *Proceedings of the U.S. National Academy of Science* **109**(3), 717-721, DOI: 10.1073/pnas.1118525109.

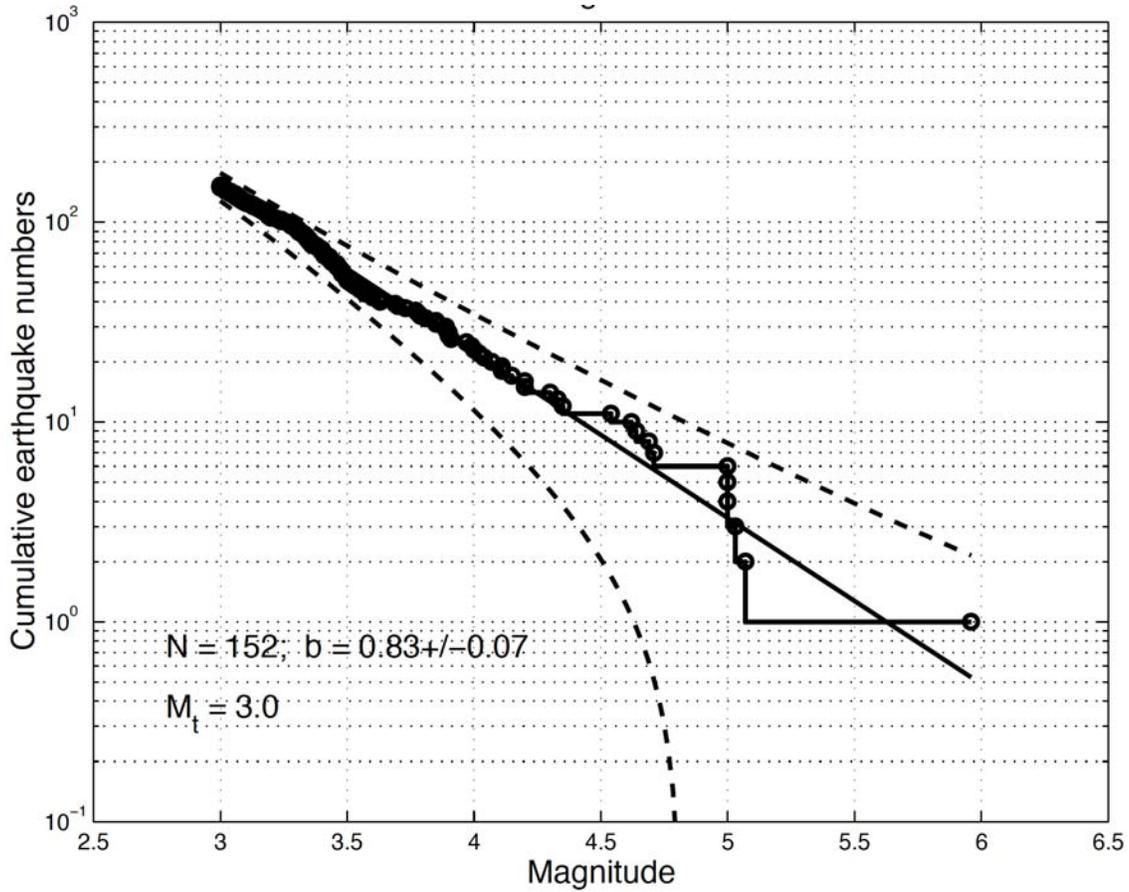

Figure 1. Magnitude–frequency relation for earthquakes from 1967 through 2005 in the Parkfield box proposed by Michael and Jones (1998). Solid line is the best fit Gutenberg-Richter approximation; dashed lines are 95% confidence limits based on Poisson occurrence. The observations fall within the range expected for a Gutenberg-Richter distribution, contrary to the characteristic model, which would imply a significant surplus of magnitude-6 "characteristic" events. (Figure adapted from Jackson and Kagan, 2006.)